\documentclass[10pt]{article}

\usepackage{amsmath}
\usepackage{amssymb}

\usepackage{graphicx}

\usepackage{cite}

\usepackage{color} 

\makeatletter
\renewcommand{\@biblabel}[1]{\quad#1.}
\makeatother

\date{}

\begin{document}

% Title must be 150 characters or less
\begin{center}
{\LARGE
\textbf{A practical recipe to fit \\ discrete power-law distributions\\}
}
\vspace{1cm}
% Insert Author names, affiliations and corresponding author email.

{\large
\'Alvaro Corral$^{1}$,
Anna Deluca$^{1,2}$, and
Ramon Ferrer-i-Cancho$^{3}$
}
\vspace{0.5cm}

$^{1}$Centre de Recerca Matem\`atica, Bellaterra, Barcelona, Spain
\\
$^{2}$Departament de Matem\`atiques, UAB, Bellaterra, Barcelona, Spain
\\
$^{3}$Departament de Llenguatges i Sistemes Inform\`atics, 
%%Universitat Polit\`ecnica de Catalunya, 
UPC, Barcelona, Spain

\end{center}

% Please keep the abstract between 250 and 300 words
\section*{Abstract}
Power laws pervade statistical physics and complex systems \cite{Bak_book,Newman_05}, 
but, traditionally,
researchers in these fields have paid little attention to properly fit these
distributions. Who has not seen (or even shown) a log-log plot of a completely
curved line pretending to be a power law?
Recently, Clauset et al. have proposed a method to decide if a set of values of
a variable has a distribution whose tail is a power law \cite{Clauset}. The key of their
procedure is the identification of the minimum value of the variable for which
the fit holds, which is selected as the value for which the Kolmogorov-Smirnov
distance between the empirical distribution and its maximum-likelihood fit is
minimum. However, it has been shown that this method can reject the power-law
hypothesis even in the case of power-law simulated data \cite{Corral_nuclear}. 
Here we propose a
simpler selection criterion, which is illustrated with the more involving case
of discrete power-law distributions.

% Please keep the Author Summary between 150 and 200 words
% Use first person. PLoS ONE authors please skip this step. 
% Author Summary not valid for PLoS ONE submissions.   

\section{Procedure}

\noindent
This method is similar in spirit to the one by Clauset et al. \cite{Clauset,Corral_nuclear},
but with important differences \cite{Peters_Deluca}. Here we just present the recipe,
the justification is available in Ref. \cite{Corral_Boleda}.
\vspace{0.5cm}

\noindent
Consider a {\bf discrete power-law} distribution, 
defined for $n=a, a+1, a+2,\dots \infty$ (with $a$ natural),
$$
f(n) = \mbox{Prob}[\mbox{variable}=n] = \frac 1 {\zeta(\beta+1,a) n^{\beta+1}}
$$
$$
S(n) = \mbox{Prob}[\mbox{variable}\ge n]=\frac{\zeta(\beta+1,n)}{\zeta(\beta+1,a)}
$$
with $\beta > 0$ and $\zeta$ the Hurwitz zeta function \cite{Clauset} (Riemann function for $a=1$),
$$
\zeta(\gamma,a) = \sum_{k=0}^\infty \frac 1 {(a+k)^\gamma}.
$$
Note then that $f(n)$ is a power law but $S(n)$ is not (only asymptotically).
\vspace{0.5cm}

\noindent
For $a$ fixed, 
the data values verifying $n\ge a$ are numbered
from $i=1$ to $N_a$, and the remainder is removed.
\vspace{0.5cm}

\newpage %%%%%%%%%%%%%%%%%%%%%%%%%%%%%%%%%%%%%%%%%%%%%%%%%%%%%%%%%%%%%%
\noindent
Then, the method consists of the following steps:

\begin{enumerate}

\item {\bf Maximum likelihood estimation} of the exponent $\beta$.

Calculate the log-likelihood function,
$$
\ell(\beta) = \frac 1 {N_a} \sum_{i=1}^{N_a} \ln f(n_i)
 = -\ln \zeta(\beta+1,a) - (\beta+1) \ln G_a,
$$
with 
%%$n_i$ the $N_a$ data in the range $n\ge a$ and 
$G_a$ the geometric mean of the data in the range, 
$\ln G_a = N_a^{-1}\sum \ln n_i $.

Calculate the maximum of $\ell(\beta)$ 
(for instance through the downhill simplex method \cite{Press}),
$$
\beta_{emp}=\max_{\forall \beta} \ell(\beta),
$$
which has an error (standard deviation \cite{Clauset}) % Alvaro: citar Newman o Clauset et al aqui???
$$
\sigma=\frac{\beta_{emp}}{\sqrt{N_a}}.
$$

The computation of the zeta function uses the Euler-Maclaurin formula \cite{Abramowitz,Vepstas},
$$
\sum_{k=0}^\infty \tilde f(k) \simeq \sum_{k=0}^{M-1} \tilde f(k) + \int_M^\infty
\tilde f(k)dk + \frac{\tilde f(M)} 2 - 
\sum_{k=1}^P \frac{B_{2k}}{(2k)!}\tilde f^{(2k-1)}(M),
$$
where $B_{2k}$ are the Bernoulli numbers 
($B_2= 1/6, B_4 = -1/30, B_6 = 1/42, B_8 = -1/30,\dots $) \cite{Abramowitz}.
So,
$$
%%\sum_{k=0}^\infty \frac 1 {(a+k)^\gamma} 
\zeta(\gamma,a)
\simeq \sum_{k=0}^{M-1} \frac 1 {(a+k)^\gamma}  
+\frac{(a+M)^{1-\gamma}}{\gamma-1}
+\frac 1 {2(a+M)^\gamma}
 +
\sum_{k=1}^P {B_{2k}}C_{2k-1}(M),
$$
with
$$
C_{2k-1}(M)=\frac {(\gamma+2k-2)(\gamma+2k-3)}{2k(2k-1)(a+M)^2} C_{2k-3}(M)
\, \mbox{ and } \, 
C_{1}(M)=\frac{\gamma}{2(a+M)^{\gamma+1}}.
$$
%%and $C_{1}(M)=\gamma /(a+M)^{\gamma+1}
The second sum in the formula runs from 
$k=1$ to a fixed $P$, taken $P=18$, except if a minimum value term (${B_{2k}}C_{2k-1}(M)$) is 
reached, case in which the sum is stopped; this ensures a better convergence \cite{Vepstas}. 
% Alvaro: deixo a les teves mans arreglar el tema de la $P$. Aquesta frase s'entendria millor sent una mica més precís. 
We also take $M=14$. 
\vspace{0.5cm}

Once we obtain $\beta_{emp}$,
how do we know if the fit is good or bad?

\item
{\bf Calculation of the Kolmogorov-Smirnov statistic} \cite{Press},
$$
d_{emp}=\mbox{max}_{\forall n\ge a}\left|\frac{N_n}{N_a}-S(n;\beta_{emp})\right|,
$$
with $N_n$ the number of data taking values larger or equal to $n$.
The maximization is performed for all values of $n \ge a$,
integer and not integer.

Large and small values of $d_{emp}$ denote respectively bad and good fits.
But what is large and small?
This is determined in Step 3.

\newpage %%%%%%%%%%%%%%%%%%%%%%%%%%%%%%%%%%%%%%%%%%%%%%%%%%%%%%%%%%%%%%

\item 
{\bf Simulation of the discrete power-law distribution}, with exponent $\beta_{emp}$
and $n\ge a$.

We use a generalization of the rejection method of Ref. \cite{Devroye}:

\begin{enumerate}
\item
Generate a uniform random number $u$ between 0 and $u_{max}$, 
with $a=1/u_{max}^{1/\beta_{emp}}$.

\item
Obtain a new random number
$$
y=\mbox{int}(1/u^{1/\beta_{emp}}),
$$
where int$(x)$ means the integer part of $x$. Notice that 
its probability function is
$$
q(y)=(a/y)^{\beta_{emp}}-(a/(y+1))^{\beta_{emp}}.
$$
%%% $= 1/\sqrt[\beta_{emp}]{u}$.

\item
Accept $y$ as the simulated value 
if a new uniform random number $v$ (between 0 and 1)
fulfills 
$$
v \le %%%\frac{f(y)}{c q(y)} = 
\frac{f(y)q(a)}{f(a)q(y)}
$$ 
and reject $y$ otherwise. If accepted, take $n=y$.\\
Notice that the computation of the $\zeta$ function is not required.

Defining  $\tau=(1+y^{-1})^{\beta_{emp}}$
and $b=(a+1)^{\beta_{emp}}$
the acceptation condition becomes simpler,
$$
v y \frac {\tau-1}{b-a^{\beta_{emp}}} \le \frac {a\tau} b,
$$

\item 
Repeat the process until $N_a$ values of $n=y$ are obtained.

\end{enumerate}

\item Apply step 1 (maximum likelihood estimation) to the simulated data.

Call the obtained exponent $\beta_{sim}$.

\item Apply step 2 (calculation of the Kolmogorov-Smirnov statistic)
to the simulated data, using the fit obtained in step 4, as
$$
d_{sim}=\mbox{max}_{\forall n\ge a}\left|\frac{N_{sim}(n)}{N_a}-S(n;\beta_{sim})\right|,
$$
with $N_{sim}(n)$ the number of simulated data taking values larger or equal to $n$.

\item 
Comparison of the 2 statistics $d_{emp}$ and $d_{sim}$ is not enough, so:

Repeat steps 3, 4, and 5 
a large enough number of times (e.g., 100 or more, as allowed by computational resources),
in order to get an ensemble of values of $d_{sim}$.

\item
Compute $p-$value as
$$
p=\frac {\mbox{number of simulations with } d_{sim}> d_{emp}} 
{\mbox{number of simulations}}.
$$
% Alvaro: cal donar l'error del p-value? es important per al poster??? Ho dic per guanyar espai.
The error of the $p-$value comes from that of a binomial distribution,
$$
\sigma_p =\sqrt{\frac{p(1-p)}{\mbox{number of simulations}}}.
$$
Low values of $p$, like $p \le 0.05$ are considered bad fits.\\
For higher values, $p>0.05$, the power-law fit with $\beta_{emp}$ cannot be rejected.

\end{enumerate}

% Alvaro: jo explicaria millor com s'explora $a$ (breument)
\noindent
Repeating the whole procedure for ``all'' values of $a$
we obtain a set of acceptable pairs of $a$ and $\beta_{emp}$.\\
Select the one that gives the smallest value of $a$
provided that $p$ is above 0.20 (for instance). In a formula,
$$
a^*=\min\{a \mbox{ such that } p > 0.20 \},
$$
which has associated the resulting exponent $\beta_{emp}^*$.

\vspace{0.5cm}
\noindent
Note that the final $p-$value of the procedure is not the
one obtained for fixed $a$, but this is not relevant 
in order to provide a good fit (as long as the latter is larger
than, say, 0.20).

\vspace{0.5cm}
\noindent
The figures illustrate the results for 
$n=$ word frequencies in the Finnish novel 
{\it Seitsem\"{a}n veljest\"{a}} %% (1870) 
by Aleksis Kivi,
for which $a^*=1$ and $\beta_{emp}^*=1.13 \pm 0.01$,
with $N_{a^*}=22035$ and $8.1 \times 10^4$ word tokens.  
Notice that $f(n)$ is a power law but $S(n)$ is not,
but both are representations of a power-law distribution.

\vspace{0.5cm}
\noindent
We thank 
R. D. Malmgren (for discussions), L. Devroye (for his book),
G. Boleda (for many things!),
and the assistants to the 2012 meeting of the network {\tt complexitat.cat}
(for their interest).

%%\bibliographystyle{plain}  % alfabetico
%%%%\bibliographystyle{apalike}
%%\bibliographystyle{unsrt}   % por orden de cita
%%\bibliographystyle{nature}

%\bibliography{../../biblio}
%%%\bibliography{biblio}
%%%\bibliography{ramon}

\begin{figure}[!ht]
\begin{center}
\vspace{-2cm}
\includegraphics[width=5in]{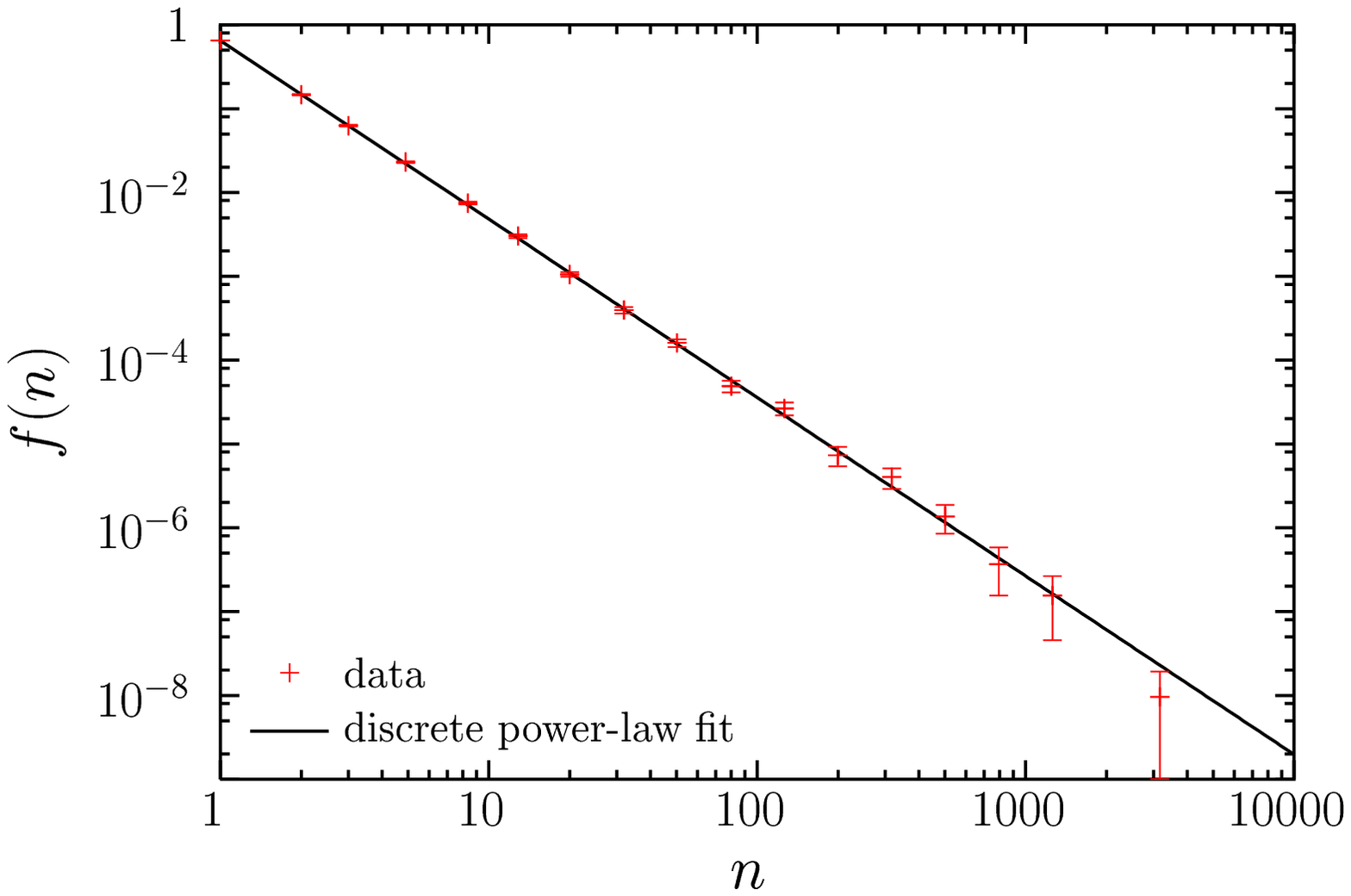}\\
\vspace{-0.5cm}
\includegraphics[width=5in]{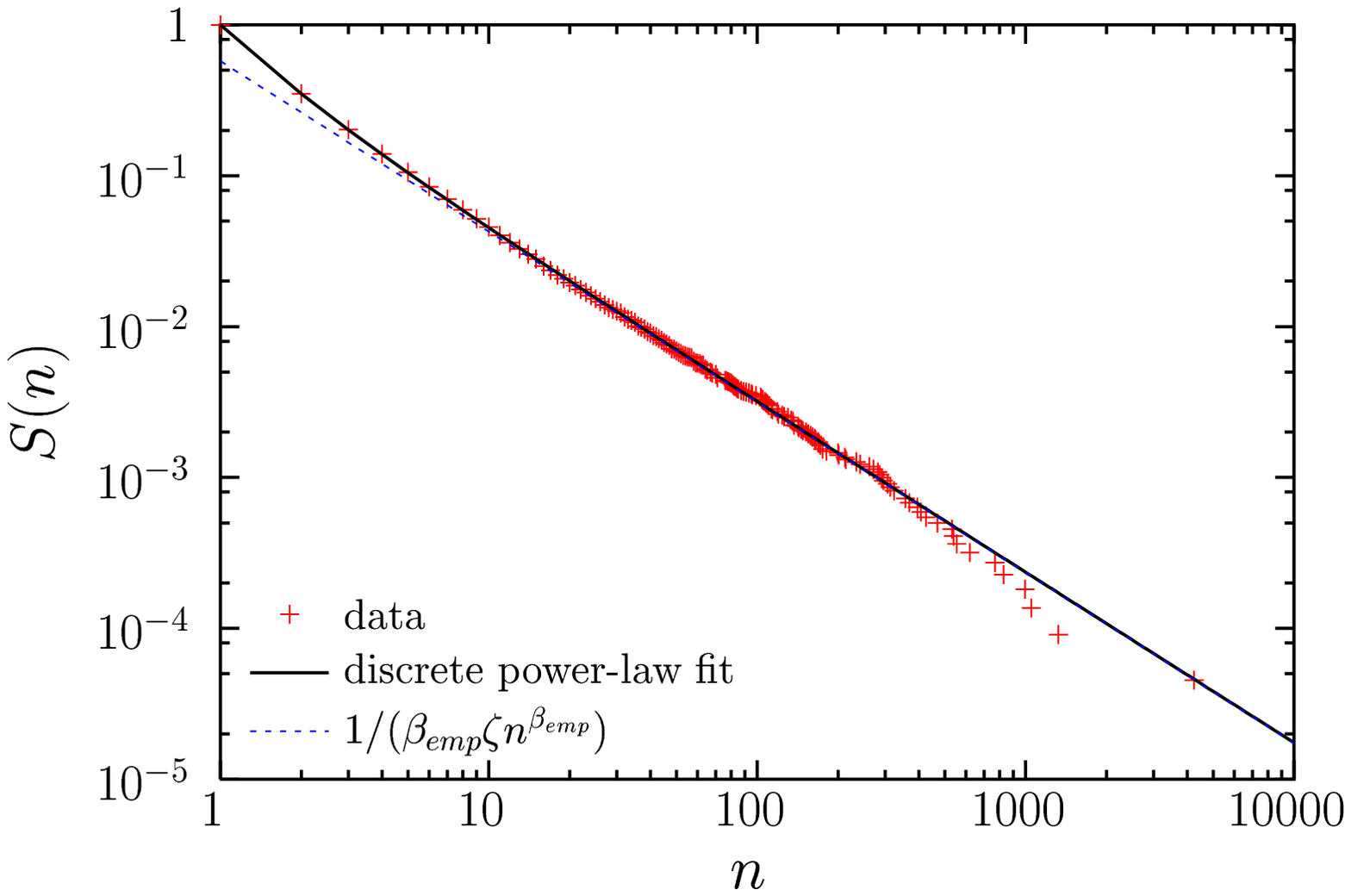}\\
\vspace{-0.5cm}
\includegraphics[width=5in]{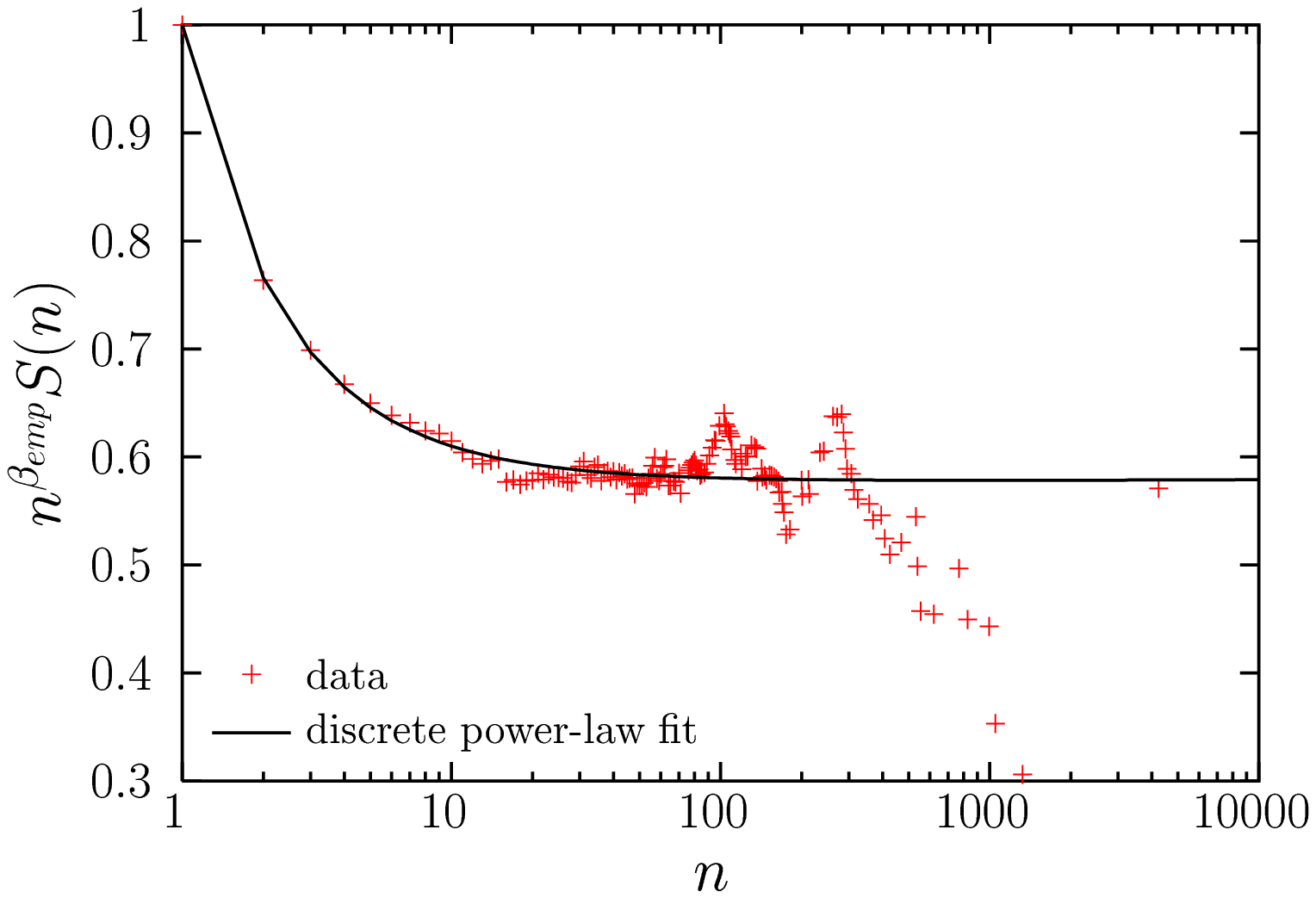}\\
\end{center}
%\caption{}
\label{Fig1a}
\end{figure}

\end{document}